# Isothermal magnetic entropy behavior in $Tb_5Si_3$: Sign reversal and non-monotonic variation with temperature and implications

Niharika Mohapatra[a, b, *], Sitikantha D Das[b,1], K. Mukherjee[b] and E. V. Sampathkumaran[b]

[a]Indian Institute of Technology Bhubaneswar, Samantapuri, Bhubaneswar, 751013, India

[b]Tata Institute of Fundamental Research, Homi Bhabha Road, Colaba, Mumbai, 400005, India

**Abstract**

The magnetic entropy change ($\Delta S$), a measure of the magnetocaloric effect, in $Tb_5Si_3$, a compound exhibiting unusual positive magnetoresistance following a magnetic-field-induced transition below magnetic transition temperature (~ 69 K), has been investigated. We found that $\Delta S$ is negative in the paramagnetic state. At the magnetic transition temperature, $\Delta S$ shows sign reversal from negative (in the paramagnetic state) to positive value in the magnetically ordered state. The high-field state which is interestingly the high resistive state is found to be associated with higher entropy i.e. large positive $\Delta S$, behaving like a paramagnet. On the basis of this observation, we conclude that the magnetic field induces magnetic fluctuations in the system resulting in positive magnetoresistance, thereby rendering support to the idea of 'inverse metamagnetism' in this compound. In addition, we note that Arrott plots present an interesting scenario.

## 1. Introduction

The rare earth based intemetallics and oxides have attracted an upsurge of research activities in the field of condensed matter physics owing to their multifunctional properties such as giant magnetocaloric effect, giant magnetoresistance, shape memory effect, magnetoelastic effect and multiferroic properties etc [1-6]. The key factor driving such multifunctinality is the coupling of magnetic subsystem with other degrees of freedom which often results from field-induced metamagnetic transition, accompanying sharp changes in the physical properties such as magnetization, magnetoresistance (MR), heat capacity, lattice parameter etc. Therefore the metamagnetic systems have recently been exploited for applications in magnetic refrigerators [1], ultrafast magnetic switching [7], memory and spintronic devices [3, 8] etc and are considered as new technological materials with tremendous potential.

We have recently reported the unusual behavior of a binary intemetallic, $Tb_5Si_3$, exhibiting magnetic field-induced transitions [9, 10]. The noteworthy behavior of this compound was the observation of a huge positive MR while increasing magnetic field beyond a critical field. The low-field state of $Tb_5Si_3$ can be regained by a field-cycling. This is interesting as a field induced transition in an antiferromagnet generally transforms a high resistive state to a low resistive state resulting in giant negative MR in the vicinity of the metamagnetic transition [2-5]. The observation of huge positive MR in the present case prompted us to do further study on $Tb_5Si_3$ to understand the origin of such a behavior [10]. In an analogy to the magnetoresistance behavior of Y doped $ErCo_2$ [11], we proposed that the application of magnetic field leads to spin-disorder (fluctuation) in the system, a phenomenon termed as "Inverse metamagnetism". Here, we report here the isothermal magnetic entropy change ($\Delta S = S(H) – S(0)$) as a function of field and temperature, derived from magnetization (*M*) data, to obtain a support to this picture.

## 2. Experimental Details

The polycrystalline sample of $Tb_5Si_3$ was prepared by conventional arc melting as described in Ref. 9. The detailed characterization techniques are also given in the same report. Further magnetic measurements were performed, with the help of a commercial superconducting quantum interference device (SQUID) magnetometer (Quantum Design) and a commercial vibrating sample magnetometer (Oxford Instruments).

## 3. Results and Discussions

The compound $Tb_5Si_3$ undergoes a disorder to order transition below ~ 69 K ($T_N$), which has been established by the anomalies in magnetization, specific heat and electrical resistivity data [9, 10]. In the present investigation, we have measured the isothermal magnetization, *M(H)*, in the temperature range 1.8 – 140 K for applied fields up to 70 kOe to evaluate isothermal *ΔS*. The temperature interval for these measurements was chosen as 5 K for (1.8 - 50 K), 2K for (50 – 70 K) and 10 K for (70 – 140 K). Selected magnetic isothermals of $Tb_5Si_3$ for the increasing field cycle are presented in Fig.1. As is evident from the figure, there is a clear distinction between magnetization behavior for the *T* ranges above and below $T_N$. While *M(H)* is found to be essentially linear with *H* without noticeable hysteresis above $T_N$, a magnetic transition at 1.8 K is apparent by the sharp change in *M(H)* in the vicinity of 60 kOe as already discussed in Ref. 9. However this transition is smoothened in the intermediate temperature range $T_N$ to 25 K. The critical field ($H_{cr}$) that induces the transition decreases with increasing temperature. $H_{cr}$, estimated from the maximum in the *dM/dT* vs. *H* plot, is shown in the inset of Fig.1. This trend observed for $H_{cr}$ is in contrast to the behavior generally seen in $RCo_2$ compounds in which the critical field increases with increasing temperature [6]. On the other hand, a similar behavior of $H_{cr}$ has been reported for the compounds such as $Nd_7Rh_3$ [2], $Nd_6Co_{1.67}Si_3$ [12], $CeFe_2$ [5], manganites [4] etc. This difference in the behavior of $H_{cr}$ may be due to the fact that the metamagnetic transition in the former set of compounds is paramagnetic to ferromagnetic (FM) state while it is antiferromagnetic (AFM) to FM state in the latter set of compounds.

The temperature variation of the critical field is found to be exponential in nature with the functional form, [$H_{cr}$ = 51.765 – 0.56 exp [T/16.19]] (shown by the dotted curve in the inset of Fig.1). But below 25 K (close to tricritical point, Ref. 9), there is a deviation of $H_{cr}$ from exponential behavior, which implies that the nature of the phase transition is different below 25 K. However a careful look at the Arrott plots (log – log $M^2$ vs. *H/M* plot), shown in Fig.2, reveals interesting features. It has been reported that negative slope or S-shaped curve is a signature of first order phase transition [6, 13] while usually observed positive slope refers to second-order magnetic phase transition. However it is fascinating to note that negative slopes are observed for $Tb_5Si_3$ in the entire temperature range below $T_N$. Even near $T_N$, the initial slope appears to be negative. There is no qualitative change in the nature of the curves around the tricritical point. These findings are puzzling to us at the moment.

We now focus on the behavior of *M(H)* in the high field state. The notable feature is that there is no complete saturation of magnetization following the metamagnetic transition. *M(H)* increases with further increasing *H* beyond $H_{cr}$ as though paramagnetic fluctuations are superimposed over ferromagnetic background. This indicates that the high field state is not a fully ferromagnetically ordered state. As magnetocaloric effect (MCE) is useful to obtain the evidence of spin fluctuation/disorder in the system, it is pertinent to investigate the field-induced change in *ΔS*.

The isothermal entropy change is determined from *M(H)* data by using the relationships derived from the Maxwell's equation as follows:

$$\Delta S = S(T,H) - S(T,0) = \int_0^H \left[\frac{\partial M(T,H)}{\partial T}\right]_H dH$$

The results of *ΔS* thus obtained from the above relationship as a function of *T* is shown in Fig. 3. We found that *ΔS* is negative in the paramagnetic state and its magnitude increases monotonically with increasing *H* as expected due to the reduction of spin fluctuation. It shows a maximum at a temperature higher temperature than $T_N$ (~ 80 K). The interesting feature is that *ΔS* changes sign exactly at $T_N$ mimicking inverse magnetocaloric effect (MCE). Although a moderate value of *ΔS* is observed here on either side (+ve and −ve) of the magnetic transition temperature, the effective change in *ΔS* around $T_N$ is 8 J/mol K for a field change of 50 kOe, comparable to the magnitude observed in many of the magneto-caloric compounds [1]. *In the figure, we also note that the peak value of ΔS at a given temperature just below $T_N$ increases with increasing H up to 40 kOe and then decreases for further increment in H.* At low temperatures (below 25 K), *ΔS* shows another peak near 7 K for *H* > 50 kOe. It may be recalled that the tricritical temperature falls in the vicinity 25 K for this compound [9] and it is possible that the applicability of Maxwell's equation to obtain *ΔS* is questionable [14] in the first-order transition regime. In order to get a better picture, the *H*-dependence of *ΔS* is plotted in Fig. 4. At the lowest observable temperature, i.e. at 3.4 K, *ΔS* exhibits a sharp jump at $H_{cr}$ (= 60 kOe) very much similar to the behavior of *M(H)* and MR near this temperature. A correlation of *M(H)*, MR and $\Delta S$ [15] is demonstrated in Fig. 5, in which we have shown the influence of the magnetic field on these properties. Magnetoresistance generally have the contributions from the change in electronic orbit due to Lorentz force and the change in electron scattering due to spin fluctuation. The Lorentz force was ruled out in Ref. 9. Therefore the positive MR arises from the spin fluctuations only which is also reflected in the $\Delta S$ behavior. The same trend of $\Delta S$ has been observed at all temperatures below 25 K while the shape of $\Delta S$ curve is different for $T > 25$ K. Looking at figure 5, it appears that $\Delta S$ tends to increase beyond the critical field, whereas MR decreases. This may be attributed to the fact that the magnetization probes the bulk of the sample, whereas electrical resistivity might respond differently to local inhomogeneities in a subtle way. The main point of emphasis is that $\Delta S$ gets even more positive following the field-induced transition, even in the temperature regime where the transition is second-order, thereby implying an additional contribution of spin disorder. In the $T$ range 67 – 25 K, $\Delta S$ shows sluggish field dependence following a dramatic increase up to a characteristic field. This characteristic field coincides with $H_{cr}$, obtained from $dM/dH$ vs. $H$ curve at fixed $T$. Such a type of $\Delta S$ behavior in a metamagnetic material is rare. It is important to note that the curves of $\Delta S(H)$ above the peak in figure 4 for $T < T_N$ are parallel to those in the paramagnetic state, thereby rendering support to the idea of "inverse metamagnetism".

## 4. Conclusions

We have investigated the isothermal entropy change behavior of $Tb_5Si_3$ exhibiting an unusual positive magnetoresistance following a field induced transition. We found that $\Delta S$ is negative in the paramagnetic state and changes its sign near the AFM transition temperature. In the magnetically ordered state (even in the temperature range where the magnetic-field-induced transition is not first-order), $\Delta S$ becomes more positive beyond certain magnetic fields, following the trend in MR behavior reported earlier [9]. But it is found to vary non-monotonically with $H$ below $T_N$ and *the curves interestingly are parallel to those of the paramagnetic state.* This provides a support to the idea of inverse metamagnetism [10]. Another important finding is that the shapes of the Arrott plots are quite puzzling.

*Corresponding Author : niharika@iitbbs.ac.in
Present address: Cavendish Laboratory, University of Cambridge, UK

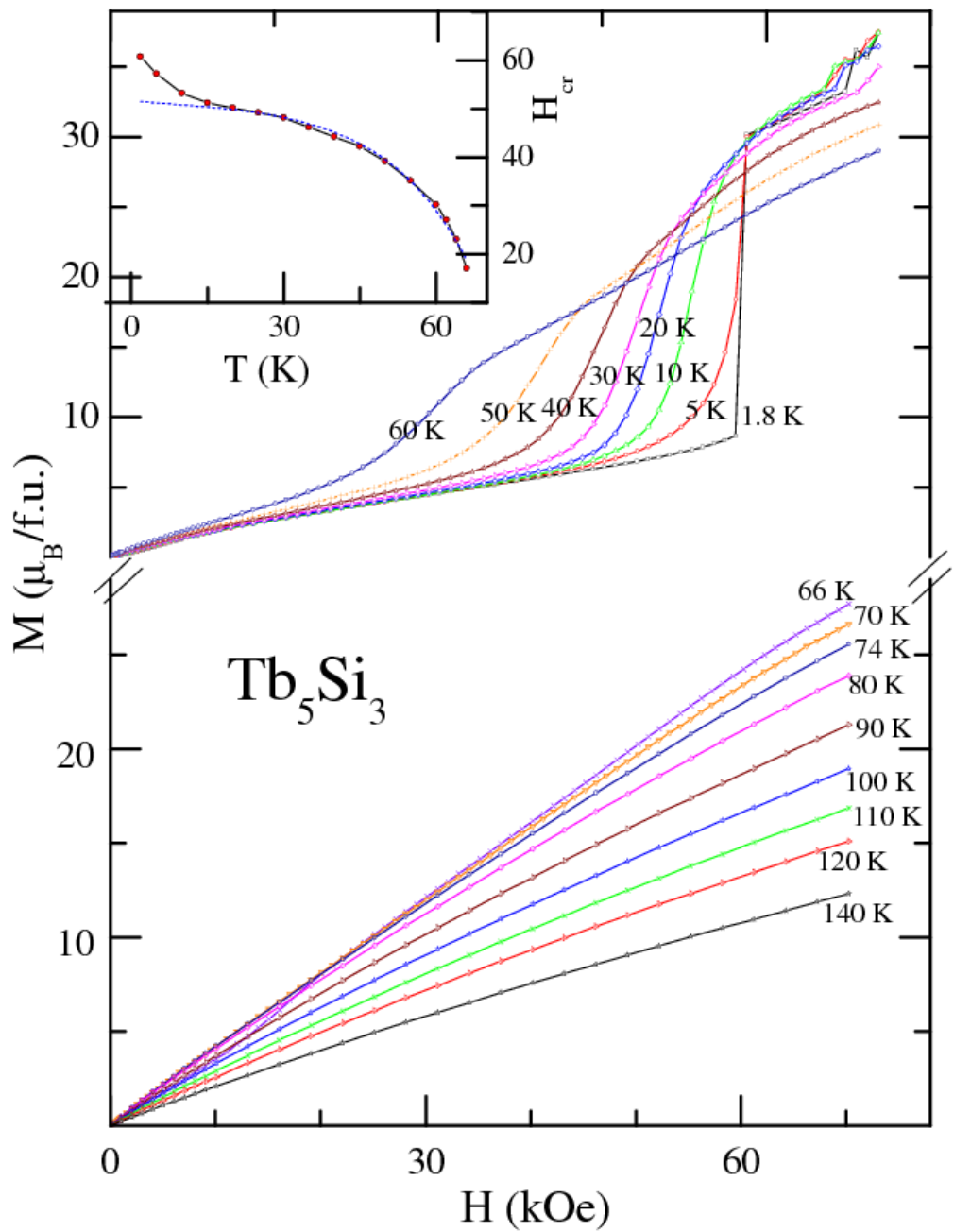

**Fig. 1.**: (color online) Magnetization isotherms of $Tb_5Si_3$ at selected temperatures measured on increasing magnetic fields up to 70 kOe. Inset shows the critical field ($H_{cr}$) as a function of temperature. The dashed line in the inset shows exponential fit to the data.

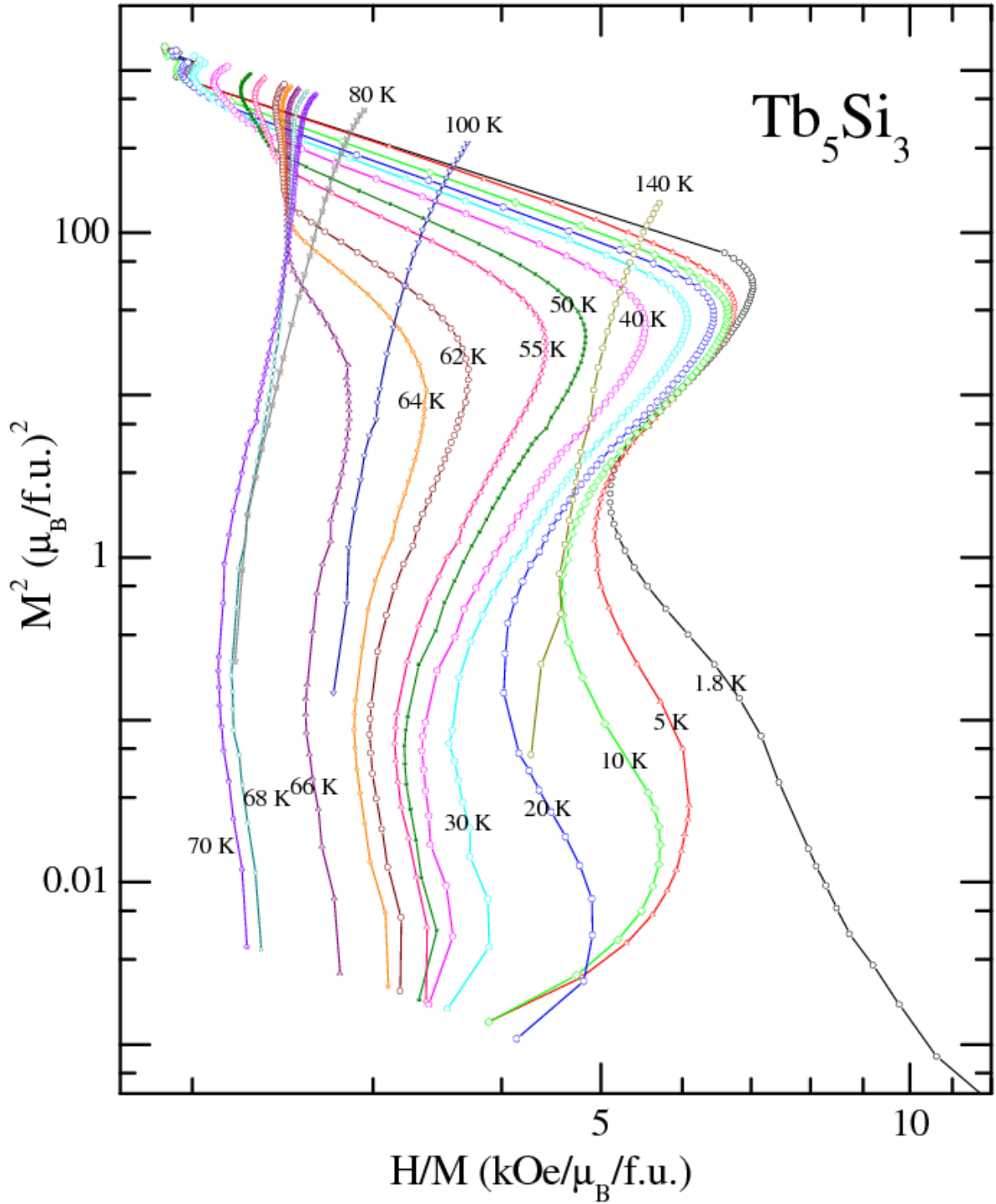

**Fig. 2.**: (color online) Arrott plots (log – log plot of $M^2$ vs. H/M) of $Tb_5Si_3$ at selected temperatures.

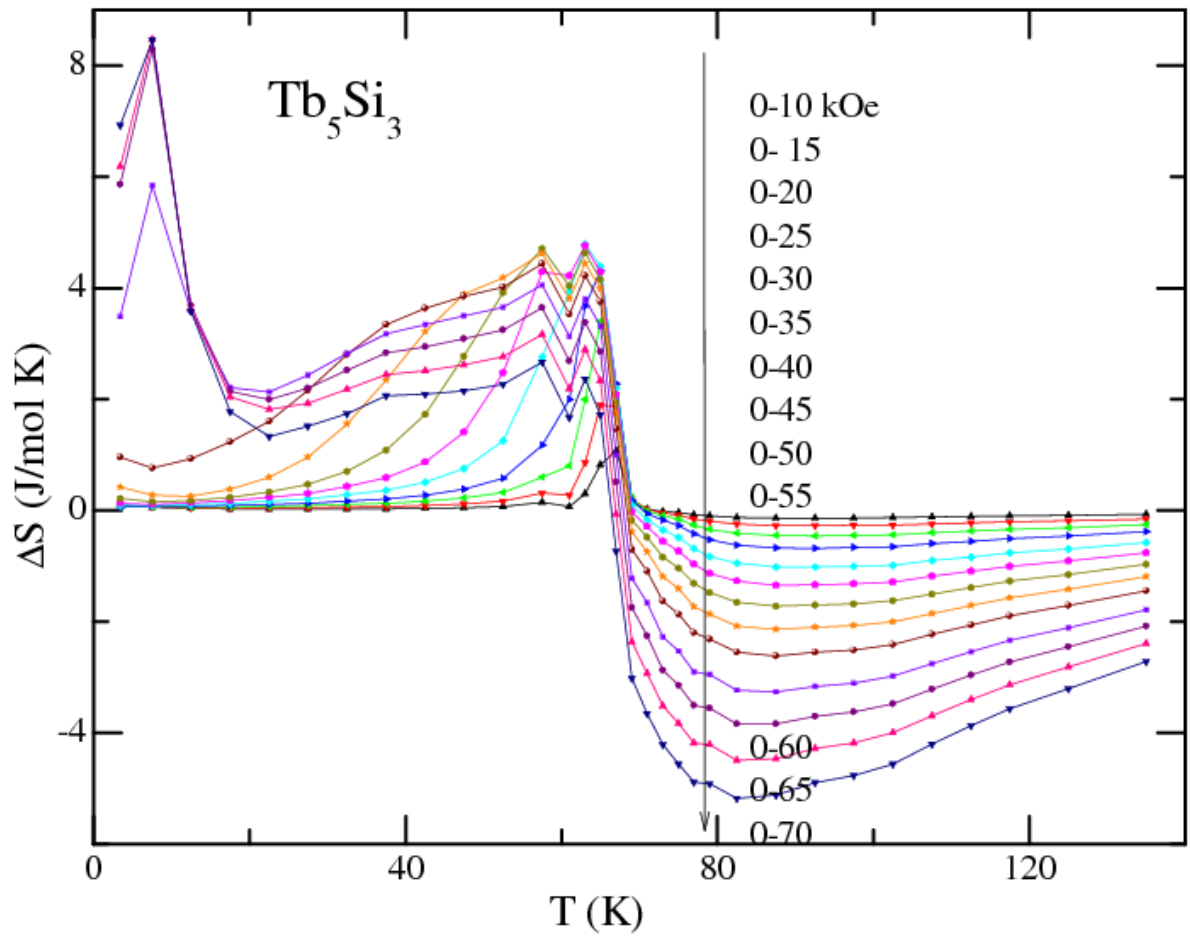

**Fig. 3.**: (color online) Isothermal entropy change (ΔS) of Tb$_5$Si$_3$ as a function of temperature for magnetic fields up to 70 kOe.

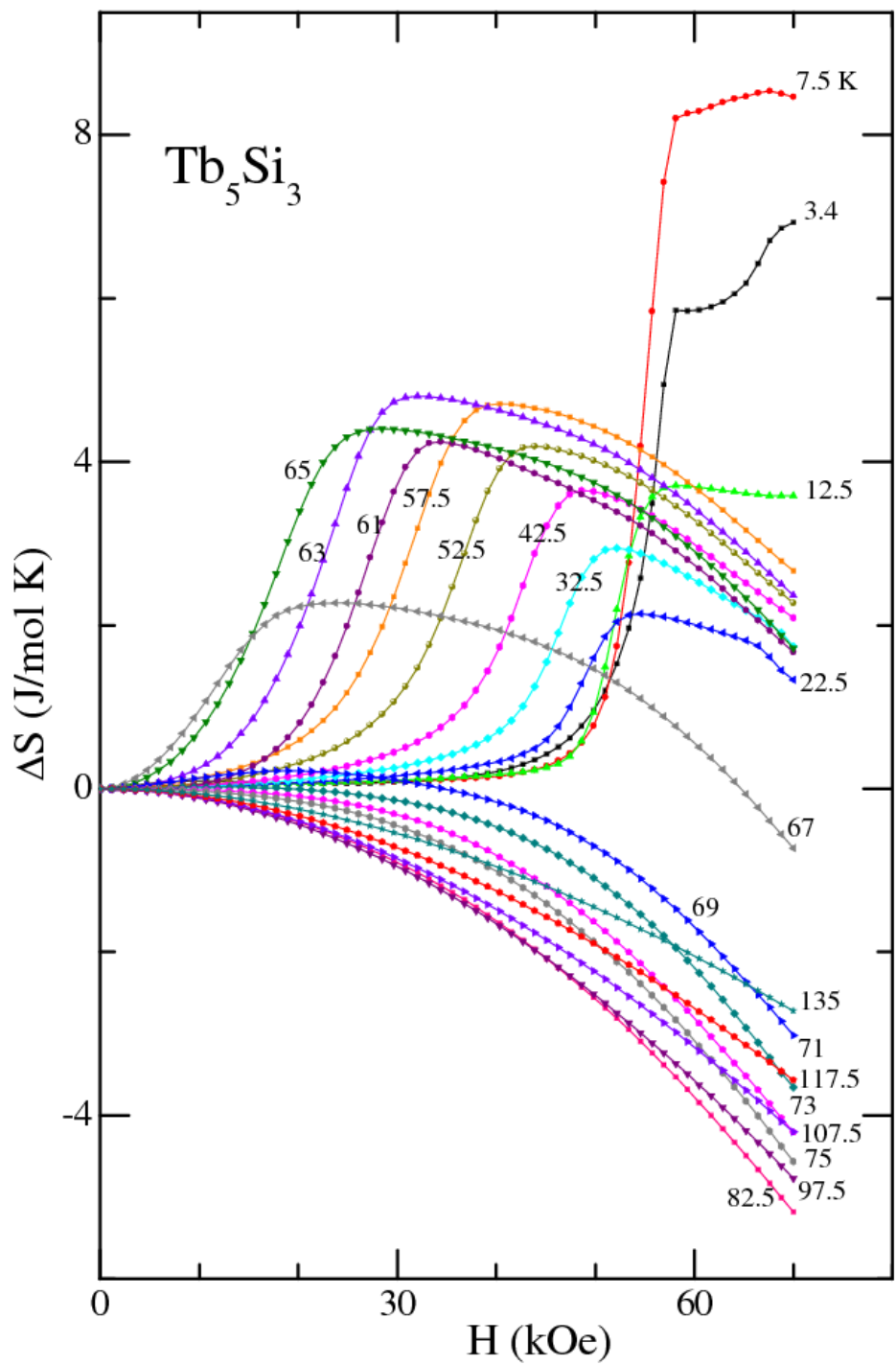

**Fig. 4.**: (color online) Isothermal entropy change (ΔS) of Tb$_5$Si$_3$ as a function of magnetic fields at selected temperatures.

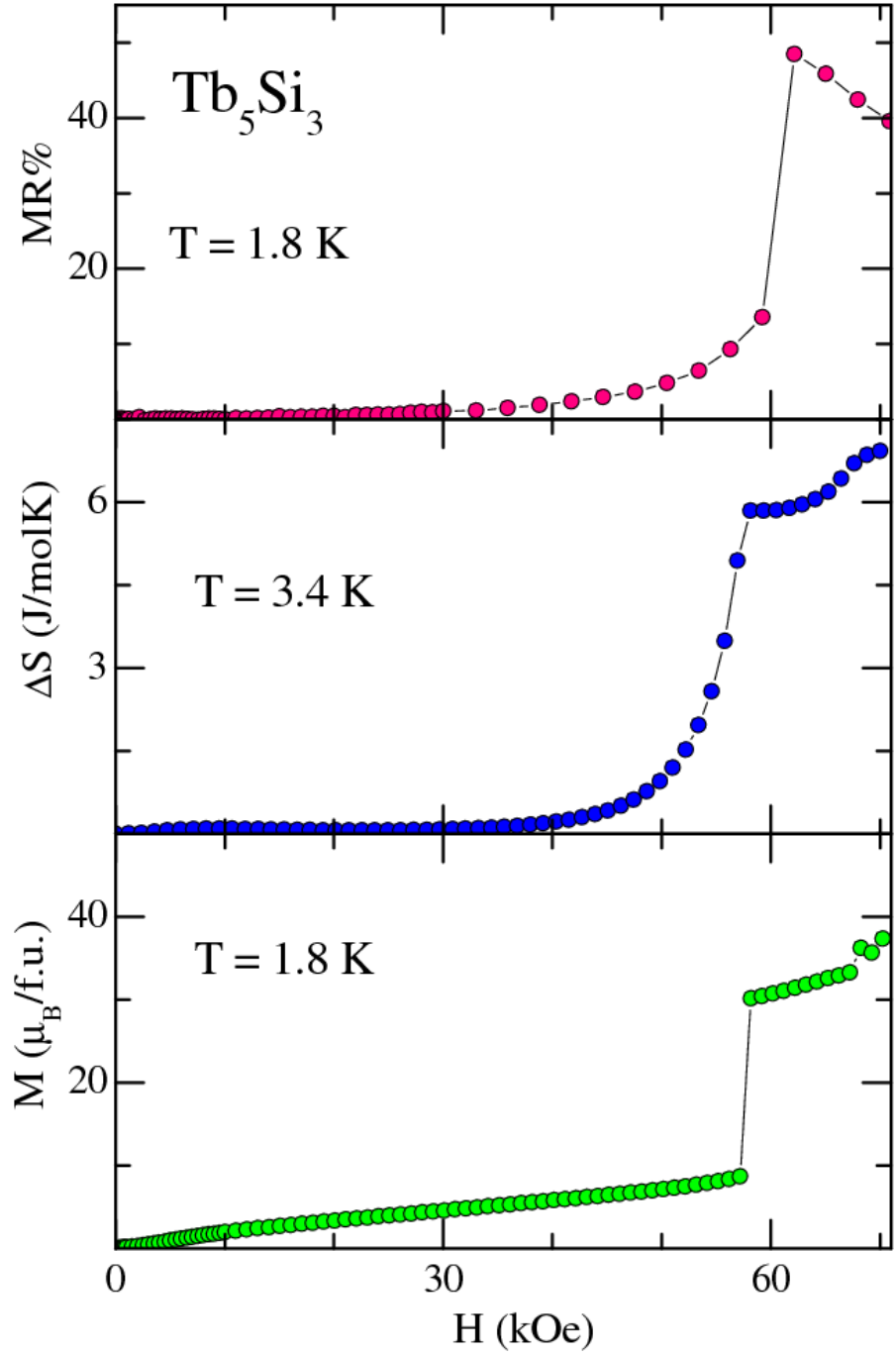

**Fig. 5.**: (color online) Magnetization, magnetoresistance and magnetic entropy change as a function of magnetic field.